\documentclass{osa-article}

\journal{osajournal}

\usepackage{lineno}
\usepackage{bm}
\usepackage{adjustbox}
\usepackage{multirow}

\articletype{article}
\begin{document}

\title{Integrated microheater on the 4H-silicon-carbide-on-insulator platform and its applications}
\author{Wenhan Sun\authormark{+1}, Ruixuan Wang\authormark{+1}, Jingwei Li\authormark{1}, Haipeng Zhang\authormark{2}, Zhensheng Jia\authormark{2}, and Qing Li\authormark{1*}}

\address{$^{1}$Department of Electrical and Computer Engineering, Carnegie Mellon University, Pittsburgh, PA 15213, USA}
\address{$^{2}$CableLabs, 858 Coal Creek Circle, Louisville, Colorado 80027, USA.}

\address{\authormark{+} These authors contributed equally to this work.}
\email{\authormark{*}qingli2@andrew.cmu.edu} 

\begin{abstract}
 Recent progress in the 4H-silicon-carbide-on-insulator (4H-SiCOI) platform has resulted in the demonstration of essential building blocks such as low-loss waveguides and microresonators. In this work, we add tunability to the 4H-SiCOI platform by integrating microheaters with compact microresonators. The strong thermo-optic effect in SiC enables a resonance tuning rate of $11.7$ pm/mW for a 36-$\mu$m-radius SiC microring, with a maximum wavelength shift up to $2.4$ nm (300 GHz). The thermal time constant of the microheater is estimated near 7 $\mu$s, corresponding to a 3-dB electrical bandwidth of 40 kHz. As a demonstration of potential applications, we employ the microheater to perform fast thermo-optic scans to deterministically access the single-soliton state of a 36-$\mu$m-radius microcomb source. In addition, an add-drop filter based on an over-coupled 18-$\mu$m-radius SiC microring is combined with the microcomb source for the selective filtering of individual comb lines, featuring an approximate 3-dB bandwidth of 5 GHz and an insertion loss of less than 1 dB. With such demonstrations, our work brings the much-needed tunability and reconfigurability to the 4H-SiCOI platform, paving the way for a wealth of chip-scale applications. 
\end{abstract}

\section{Introduction}
The 4H-silicon-carbide-on-insulator (SiCOI) photonic platform has attracted considerable research interests due to its unique advantages for photonic and quantum applications \cite{Vuckovic_SiC_review}. For example, low-loss waveguides and microresonators demonstrated in the 4H-SiCOI platform have already attained a propagation loss of $<0.1$ dB/cm \cite{Noda_4HSiC_PhC, Vuckovic_4HSiC_nphoton, Ou_4HSiC_combQ}. In addition, its strong third-order nonlinear coefficient (up to four times of that of silicon nitride) enables power-efficient four-wave mixing, which underpins critical applications such as Kerr microcombs and entangled photon pair generation \cite{Vuckovic_4HSiC_soliton, OuXin_soliton, Li_4HSiC_comb, Li_SiC_entangled}. Moreover, discrete and strong Raman and Brillouin shifts possessed by 4H-SiC, thanks to its single-crystal nature, have been exploited for chip-scale Raman and Brillouin lasers \cite{Ou_4HSiC_combQ, Li_SiC_Raman, Li_4HSiC_SBS}. Despite these impressive progresses, to date the performance metrics of 4H-SiC building-block components are mostly static and cannot be dynamically tuned or reconfigured, hence significantly hindering their scalable deployment. Such a lack of tunability is partly attributed to the small electro-optic (EO) coefficient ($<$ 1 pm/V) found in 4H-SiC \cite{Li_SiC_EOM}, which necessitates  excessively large on-chip voltages for modest device tuning. 

In addition to EO tuning, thermo-optic tuning is another common approach to adding tunability to photonic integrated circuits. For example, microheaters have been introduced to silicon-rich amorphous SiC \cite{aSiC_microheater} and 3C-SiCOI \cite{Adibi_3CSiC_microheater} to tune the refractive index of the materiel through local temperature variation. For 4H-SiC, only the temperature of the whole photonic chip has been utilized for thermo-optic tuning \cite{Ou_DTU_4HSiC_thermal}, which cannot be extended to complicated photonic circuitry. In this work, we add tunability to the 4H-SiCOI platform by integrating power-efficient microheaters with compact microresonators. For example, we report a $11.7$ pm/mW resonance tuning rate for a 36-$\mu$m-radius SiC microring, and employ the fast thermo-optic scans to deterministically access the single-soliton state for a fixed laser wavelength. In addition, an add-drop filter based on over-coupled 18-$\mu$m-radius SiC microring is also demonstrated, which exhibits a tuning rate of $21.6$ pm/mW and a maximum tuning range of 300 GHz. Such a high-performance add-drop filter is combined with the 36-$\mu$m-radius microcomb source to showcase selective filtering of individual comb lines, featuring an approximate 3-dB bandwidth of 5 GHz and an insertion loss of less than 1 dB.   

\section{Microheater design, fabrication and characterization} 
\begin{figure}[ht]
\centering
\includegraphics[width=0.65\linewidth]{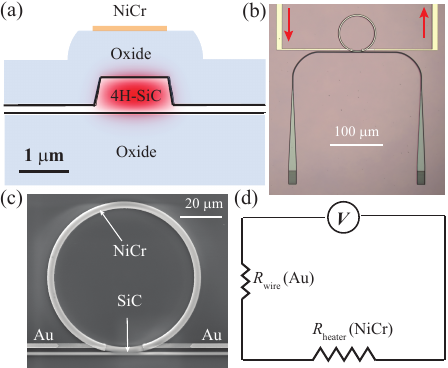}
\caption{(a) Schematic of an integrated nichrome (NiCr) microheater on a 4H-SiC rib waveguide. (b) Optical micrograph of a grating-coupled SiC microring resonator integrated with a microheater. The red arrows signal the current flow through the microheater. (c) An scanning electron micrograph of a 36-$\mu$m-radius microring with the microheater implemented on the top surface. (d) Resistive heating model of the microheater. $R_{\mathrm{heater}}$: NiCr heater resistance; $R_{\mathrm{wire}}$: Au wire resistance; and $V$: applied voltage.}
\label{Fig_schematic}
\end{figure}
\noindent As illustrated in Fig.~\ref{Fig_schematic}(a), the microheater functions by delivering resistive heating power to the resonator region for local temperature variation. Therefore, it is desirable to maximize the resistivity contrast between the heater material and the connecting wire (see Fig.~\ref{Fig_schematic}(b)). In this work,  we choose nicrhome (NiCr) and gold (Au) for the heater and wire, respectively (see Fig.~\ref{Fig_schematic}(c) and \ref{Fig_schematic}(d)). This is because NiCr has a much higher resistivity compared to Au and is also robust to elevated temperatures induced by a large current. In addition, NiCr is resistant to corrosion due to its natural chromium dioxide coating when heated and exposed to air.

For nanofabrication, a 4H-SiCOI chip (NGK Insulators) consisting of a $700\text{-}\mathrm{nm}$-thick SiC layer on top of a $2\text{-}\mathrm{\mu m}$-thick silicon dioxide ($\mathrm{SiO_2}$) is employed. We first pattern the optical layer using flowable oxide (FOx-16) with e-beam lithography (EBL), which is followed by fluorine-based dry etching for the removal of approximately $600\;\mathrm{nm}$ of the SiC layer while leaving a $100\text{-}\mathrm{nm}$-thick pedestal \cite{Li_4HSiC_comb}. Next, we proceed to fabricate the microheater and its accessories (i.e., wires and electrode pads) based on a liftoff process. For this purpose, a $1\text{-}\mathrm{\mu m}$-thick $\mathrm{SiO}_2$ layer is deposited on top of SiC using plasma-enhanced chemical-vapor deposition (PECVD), serving as a cladding layer between the SiC microresonator and the microheater. After that, a $1\text{-}\mathrm{\mu m}$-thick PMMA 950 A7 resist is patterned in EBL with the help of alignment markers, so that the microheater is aligned to the underlying microring resonator. We then sputter a $100\text{-}\mathrm{nm}$-thick NiCr layer, on top of which an additional $40\text{-}\mathrm{nm}$-thick gold is deposited to reduce the overall resistance. Metal layers that are not part of the electrical circuit are then lifted off. Finally, an opening window surrounding the microresonator is defined using EBL (PMMA), within which the gold layer is cleared using standard gold etchant. In Fig.~\ref{Fig_schematic}(c), one example of the fabricated microheater for a 36-$\mu$m-radius SiC microring is provided. As can be seen, the NiCr microheater encompasses the majority of the microring area, with the positioning error estimated to be less than $50$ nm.

\begin{figure}[ht]
\centering
\includegraphics[width=0.75\linewidth]{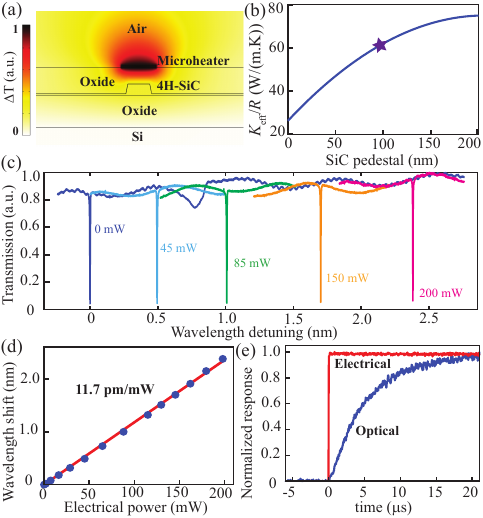}
\caption{(a) Temperature distribution induced by the microheater on top of a 4H-SiC rib waveguide based on the finite element method (FEM). (b) Simulated effective thermal conductance of the SiC microring normalized by its radius as a function of the pedestal thickness. The star corresponds to our current design with a pedestal thickness of $100\;\mathrm{nm}$; (c) Superimposed linear transmission scans of a 36-$\mu$m-radius SiC microring at various microheater powers. (d) Summarized resonance shifts (blue markers) as a function of the microheater electrical power. The linear fitting (red line) points to a thermal efficiency of $11.7\;\mathrm{pm/mW}$. (e) The optical signal (blue) in response to a sudden increase in the microheater power (red), revealing a thermal time constant around  $7\;\mathrm{\mu s}$.}
\label{Fig_DCAC}
\end{figure}

Figure \ref{Fig_DCAC}(a) displays the simulated temperature distribution induced by the microheater based on the finite element method (FEM), revealing an almost uniform temperature inside the SiC core. This result is not surprising given that the thermal conductivity of SiC is two orders of magnitude higher than that of SiO$_2$. In addition, considering that the waveguide mode is highly confined within the SiC core and that the thermo-optic efficient of SiC is several times (>5x) of that of the surrounding SiO$_2$, we can focus on the index change of SiC while ignoring that of SiO$_2$ as a good approximation. Consequently, the resonance wavelength shift induced by the microheater can be estimated as \cite{Li_SiC_EOM}:
\begin{equation}
    \frac{\mathrm{d}\lambda_r}{\mathrm{d}P} \approx \Gamma_c\frac{\lambda_r}{n_c}\frac{\mathrm{d}n_c}{\mathrm{d}T_c}\frac{\mathrm{d}T_c}{\mathrm{d}P},
    \label{Eq1}
\end{equation}
where $\Gamma_c$ is the mode confinement factor, i.e. the portion of the energy stored within the waveguide core ($\Gamma_c \approx 1$), $n_c$ is the refractive index of the waveguide core ($n_c \approx 2.6$ for 4H-SiC), $\mathrm{d}n_c/\mathrm{d}T_c$ is the thermo-optic coefficient ($\approx 4.5\times 10^{-5}$/K) \cite{Ou_DTU_4HSiC_thermal}, and $\mathrm{d}T_c/\mathrm{d}P$ is the differential slope of the core temperature's response to the the microheater power. By defining an effective thermal conductance $K_\text{eff}$ as $K_\text{eff}\equiv \mathrm{d}P/\mathrm{d}T_c$ and assuming a linear thermo-optic effect, we can further simplify Eq.~\ref{Eq1} as 
\begin{equation}
    \frac{\mathrm{d}\lambda_r}{\mathrm{d}P}\approx \frac{26.8\ \text{pm}/K}{K_\text{eff}}.  
    \label{Eq2}
\end{equation}

In Eq.~\ref{Eq2}, $K_\text{eff}$ can be directly computed based on its definition using FEM shown in Fig.~\ref{Fig_DCAC}(a). Numerical simulations suggest that $K_\text{eff}$ is mostly sensitive to the SiC pedestal thickness while only showing weak dependence on other geometric parameters such as the waveguide and microheater widths. Figure \ref{Fig_DCAC}(b) plots the radius-normalized $K_\text{eff}$ as a function of the pedestal height, so that the result is applicable to SiC microrings with different sizes. In our current design, although the pedestal is only 100 nm, the large thermal conductivity of SiC has already increased the overall thermal conductance by more than two times when compared to the no-pedestal case \cite{Li_4HSiC_direct_soliton}. Substituting the numerical value of $K_\text{eff}$ into Eq.~\ref{Eq2}, we obtain a predicted thermal efficiency of $12$ pm/mW and $24$ pm/mW for 36-$\mu$m-radius and 18-$\mu$m-radius SiC microrings, respectively. If the SiC pedestal thickness is reduced to zero, the thermal efficiency of SiC microrings is further improved by a factor of $2.5$, e.g., reaching $30$ pm/mW for 36-$\mu$m-radius microrings \cite{Adibi_3CSiC_microheater}.  

In the experiment, the thermal efficiency of the microheater is characterized by tracking the resonance shift as a function of the applied electrical power \cite{Li_4HSiC_comb}. One such example is provided in Fig.\ref{Fig_DCAC}(c), which corresponds to the fundamental transverse-magnetic (TM$_{00}$) mode of a 36-$\mu$m-radius SiC microring with a waveguide width of $1.5\ \mu$m. As can be seen, consistent resonance shifts proportional to the microheater powers are confirmed. The maximum $2.4\;\mathrm{nm}$ wavelength shift, or equivalently $300\;\mathrm{GHz}$ frequency shift, is mainly limited by the maximum tolerable current density of the NiCr layer, beyond which the microheater risks burning. By fitting the measured wavelength shift versus the applied electrical power (Fig.\ref{Fig_DCAC}(d)), we extract a thermal efficiency of $11.7\;\mathrm{pm/mW}$ for the 36-$\mu$m-radius microring, which is in good agreement with the FEM simulation (see Fig.~\ref{Fig_DCAC}(b) and Eq.~\ref{Eq2}). In addition to the DC characterization, the AC response of the microheater is also quantified by applying a step function in the electrical power. Due to the finite heat capacity, the temperature of the microresonator will only gradually increase before reaching thermal equilibrium. This temperature evolution is captured by biasing the microresonator in the linear regime of the optical resonance (i.e., near the full width at the half maximum) and recording the optical transmission using a large-bandwidth oscilloscope \cite{Li_SiC_EOM}. For example, by fitting the optical response shown in Fig.~\ref{Fig_DCAC}(e), the thermal time constant of the microheater is estimated to be $7\;\mathrm{\mu s}$. This small response time translates to a relatively large electrical 3-dB bandwidth of $40\;\mathrm{kHz}$, which is sufficient for many practical applications requiring modest to fast tuning/reconfiguring speeds. 

\section{Application: Soliton microcomb generation}
\begin{figure}[ht]
\centering
\includegraphics[width=0.72\linewidth]{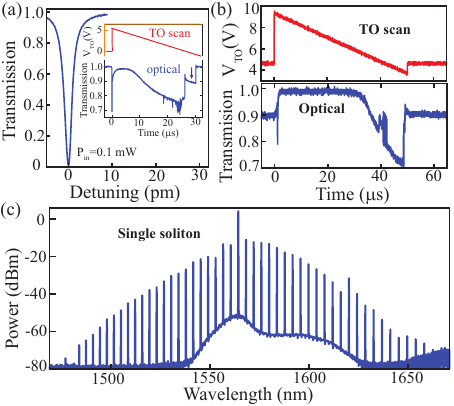}
\caption{(a) Linear swept-wavelength transmission scan of the fundamental transverse-magnetic (TM$_{00}$) resonance in a 36-$\mu$m-radius SiC microring, exhibiting an intrinsic and loaded optical quality factor of $1.4$ million and $0.7$ million, respectively. The inset plots the recorded transmission signal (lower panel) as a function of the applied voltage (upper panel) to the microheater for a fixed laser detuning and an on-chip power of 40 mW. The arrow marks the single-soliton step. (b) Optimized thermo-optic scan that repeatably generates the soliton microcomb (upper panel) and the corresponding optical transmission (lower panel), ending in the single-soliton state. (c) The corresponding optical spectrum of the single soliton microcomb generated by the thermo-optic scan in (b). }
\label{Fig_soliton}
\end{figure}

In this section, we apply the developed microheater technology for the deterministic access of the single soliton state in a SiC microresonator. A soliton microcomb is a coherent optical source implemented in nonlinear optical microresonators, whose spectrum consists of multiple equally-spaced, phase-locked individual comb lines \cite{Diddams_comb_review1}. This unique property has made the microcomb technology a transformative solution for many important applications. For example, a soliton microcomb could play an essential role in realizing efficient wavelength division multiplexing (WDM) systems, enabling ultra-high-capacity transmission over a single fiber and thus meeting the escalating bandwidth demands in modern optical networks \cite{Comb_communication}. The strong Kerr nonlinearity of 4H-SiC has rendered it an excellent material of choice for producing power-efficient soliton microcombs \cite{Vuckovic_4HSiC_soliton, OuXin_soliton, Li_4HSiC_100GHz_soliton}. However, generation of the soliton microcomb typically requires overcoming the thermo-optical bistability of the microresonator, which makes the soliton state thermally unstable. To solve this issue, various approaches have been investigated, including power kicking \cite{Kippenberg_power_kicking} and fast frequency tuning \cite{Gaeta_SiN_microheater}. For 4H-SiC microresonators, researchers have been mainly relying upon the employment of auxiliary cooling lasers \cite{OuXin_soliton, Li_4HSiC_100GHz_soliton} or slow laser frequency scanning \cite{Vuckovic_4HSiC_soliton, Li_4HSiC_direct_soliton} to create single solitons. However, both approaches have their limitations. For example, the auxiliary laser method necessitates the introduction of an additional laser to be coupled to a non-competing resonant mode, which complicates the optical design and the experimental characterization \cite{Wong_soliton_aux}. On the other hand, the approach based on the slow laser frequency scanning puts stringent requirements on the properties of the microresonator, and only a few candidates with strong Kerr nonlinearity and low enough optical loss qualify \cite{Li_4HSiC_direct_soliton, Kippenberg_SiN_100GHz_2018}. 

Here, we take advantage of the fast frequency tuning offered by an integrated microheater to deterministically access the single soliton state of SiC microresonators. This is because the thermo-optic scan not only changes the detuning of the resonant mode, but also is capable of influencing the thermal dynamics of the microresonator when scanning at fast enough speeds \cite{Gaeta_SiN_microheater}. To illustrate this process, we use the 36-$\mu$m-radius SiC microring shown in Fig.~\ref{Fig_schematic}(c) as an example. For the microring with a ring width of $1.5$ $\mu$m, the fundamental TM$_{00}$ resonance exhibits an intrinsic and loaded optical quality factor of $1.4$ million and $0.7$ million (Fig.\ref{Fig_soliton}(a)), respectively. In addition, this mode family displays strong anomalous group velocity dispersion in the 1550 nm band \cite{Li_4HSiC_comb}, thus supporting efficient microcomb generation via the Kerr effect. Indeed, if we scan the pump laser slowly across the resonance at large enough powers (>20 mW), chaotic modulation-instability (MI) combs are observed in a straightforward manner. The soliton state, however, remains inaccessible due to the thermo-optic bistability \cite{Li_4HSiC_direct_soliton}.

As shown in Fig.\ref{Fig_soliton}(a), by fixing the input laser wavelength and applying a time-varying voltage to the microheater, we observe a soliton step in the optical transmission. Note that as the microheater power decreases in the thermo-optic scan, the resonance wavelength is blue shifted (i.e., decreasing in wavelength) as the cavity temperature drops. However, even if we maintain the same scanning voltage corresponding to the soliton step (marked by an arrow in the inset of Fig.~\ref{Fig_soliton}(a)), the cavity temperature will keep dropping due to the decrease in the intracavity optical power caused by the transition from the MI comb to the soliton state, making the soliton state thermally unstable. Hence, simple thermo-optic scans as the one shown in Fig.~\ref{Fig_soliton}(a) are insufficient to stabilize the cavity temperature for the soliton generation. A more sophisticated thermo-optic scan is provided in Fig.~\ref{Fig_soliton}(b), where we add a DC bias to the scanning voltage and a slight reverse kick in the microheater power to compensate for the drop in the intracavity optical power when landing on the soliton state \cite{Gaeta_SiN_microheater}. With optimized scanning parameters such as the DC bias, voltage slope, as well as the reverse kick amplitude, the single-soliton state becomes repeatedly accessible at the end of each scan (see Fig.\ref{Fig_soliton}(b)). By triggering the microheater signal only once while maintaining a steady bias voltage afterwards, a stable soliton state is generated and maintained for a long time (>30 min). The corresponding optical spectrum shown in Fig.\ref{Fig_soliton}(c) confirms that it is a single soliton state.

\section{Application: Individual comb-line filtering}
\begin{figure}[ht]
\centering
\includegraphics[width=0.8\linewidth]{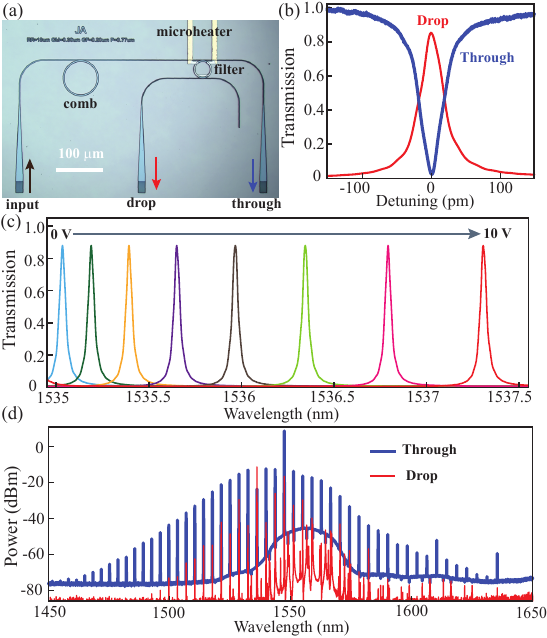}
\caption{(a) Optical micrograph of a two-microring structure for the comb line filtering: The 36-$\mu$m-radius comb ring generates the frequency comb and 18-$\mu$m-radius add-drop filter ring selects the targeted comb line and drop it to the drop port. (b) Swept-wavelength transmission scans of the filter ring resonance for the through (blue line) and drop (red line) ports. (c) Superimposed drop-port transmission scans of the filter corresponding to microheater voltages varied from $0\;\mathrm{V}$ to $10\;\mathrm{V}$. The extracted thermal efficiency is estimated to be $21.6\ \mathrm{pm/mW}$. (d) Superimposed optical spectra from the through (blue line) and drop (red line) ports when the comb ring is pumped at $1547.4$ nm while the resonance of the filter ring is adjusted to selectively filter out the comb line at $1536.1$ nm.}
\label{Fig_filter}
\end{figure}

In the second example, we employ the microheater to tune the center frequency of an add-drop filter, which is then combined with a microcomb source to demonstrate selective filtering of individual comb lines. Achieving such functionality on the chip scale is critical to a variety of practical applications of the microcomb technology, e.g., in the parallel data communication where each comb line is utilized to carry independent data \cite{Comb_communication}. As shown in Fig.~\ref{Fig_filter}(a), our device consists of two microrings with identical width of $2\ \mu$m: the first microring has a radius of 36 $\mu$m and is responsible for the microcomb generation; and the second ring has a radius of 18 $\mu$m, which is strongly over-coupled to function as an add-drop filter with its center frequency controlled by an integrated microheater. The free spectral range (FSR) of the comb and filter microrings is estimated to be 475 GHz and 963 GHz, respectively. The swept-wavelength transmission scans of the TM$_{00}$ mode in the add-drop filter are provided in Fig.~\ref{Fig_filter}(b), exhibiting an approximate 3-dB bandwidth of 5 GHz and an on-chip insertion loss of $<1$ dB. Note that for the filter ring, the loaded quality factor of the TM$_{00}$ mode ($\approx 39,000$)  is much smaller than its intrinsic quality factor ($>1$ million), confirming the strong coupling regime. Attaining such a strong level of over-coupling has two major benefits for the filter: first, the GHz-level bandwidth makes the filter less vulnerable to the environmental temperature fluctuations; and second, the relatively low quality factor also permits the filter to tolerate much higher optical powers without incurring significant nonlinear distortions. The tuning response in Fig.~\ref{Fig_soliton}(c) verifies that the center wavelength of the filter can be tuned at an approximate rate of $21.6$ pm/mW for a range of 300 GHz, consistent with our FEM simulation (see Fig.\ref{Fig_DCAC}(b) and Eq.~\ref{Eq2}). 

To demonstrate individual comb line filtering, we pump the $1547.4$ nm resonance of the comb ring with an on-chip power around 70 mW. The strong four-wave mixing results in a MI comb as shown in Fig.~\ref{Fig_filter}(d) (noting that the comb represented by the blue line is not a soliton since there is no microheater on this ring). Next, a DC voltage is applied to the microheater to align the filter's resonance to the comb line centered at $1536.1$ nm (3 FSR away from the pump). As can be seen in Fig.~\ref{Fig_filter}(d), in the through port only the $1536.1$ nm comb line experienced significant suppression ($\approx 15$ dB), while the rest comb spectra are almost the same as the case when not being filtered. This is because the FSR of the filter ring (963 GHz) is intentionally chosen to be different than twice of the comb spacing (475 GHz), as their frequency difference for every two comb lines is about 13 GHz, which is much larger than the 3-dB bandwidth of the filter ($\approx 5$ GHz). Such Vernier design ensures that within >70 THz range, only the desired comb line is filtered without accidentally affecting other lines . Finally, the optical spectrum corresponding to the drop port (red line in Fig.~\ref{Fig_filter}(d)) also confirms that the $1536.1$ nm comb line has been successfully dropped with $< 1$ dB insertion loss, while the rest comb lines are suppressed in the drop port given that they are not resonant with the filter ring.  

\section{Conclusion}
In summary, we have successfully demonstrated an integration of NiCr-based microheaters to the 4H-SiCOI platform, achieving a thermal tuning efficiency of $11.7\;\mathrm{pm/mW}$ and $21.6$ pm/mW for $36\text{-}\mathrm{\mu m}$-radius and 18-$\mu$m-radius microrings, respectively. In addition, our microheater has an approximate cutoff frequency of 40 kHz, thus enabling fast resonance tuning on the order of microseconds. We showcase potential applications by applying the fast thermo-optic tuning for the deterministic access of the single soliton state in a 36-$\mu$m-radius SiC microring. In addition, an add-drop filter based on an over-coupled 18-$\mu$m-radius SiC microring is also demonstrated, featuring a 3-dB bandwidth of 5 GHz and a maximum tuning range of $>300$ GHz. Furthermore, this add-drop filter is combined with the 36-$\mu$m-radius microcomb source to demonstrate low-loss individual comb line filtering. With these results, we believe our work has paved the way for the wide adoption of the microheater technology to 4H-SiCOI, potentially transforming a wealth of chip-scale applications.     

\begin{backmatter}
\bmsection{Funding}
This work was supported by CableLabs University Outreach Program, NSF (2240420, 2427228), as well as an award from U.S.Department of Commerce, National Institute of Standards and Technology (60NANB24D153). 

\bmsection{Acknowledgments}
The authors acknowledge the use of Bertucci Nanotechnology Laboratory at Carnegie Mellon University supported by grant BNL-78657879 and the Materials Characterization Facility supported by grant MCF-677785. R.~Wang and J.~Li also acknowledge the support of Tan Endowed Graduate Fellowship and Benjamin Garver Lamme/Westinghouse Graduate Fellowship from CMU, respectively. 

\bmsection{Disclosures}  The authors declare no conflicts of interest.

\bmsection{Data Availability} Data underlying the results presented in this paper are not publicly available at this time but may be obtained from the authors upon reasonable request.

\end{backmatter}
\bibliography{SiC_Ref}

\end{document}